\documentclass[aip,jcp,preprint]{revtex4-1}             

\usepackage{color}
\usepackage{graphicx}
\usepackage{ulem}

\begin{document}

\title{Mixing Effects in the Crystallization of Supercooled Quantum Binary Liquids}


\author{M. K\"uhnel}
\affiliation{Institut f\"ur Kernphysik, J. W. Goethe-Universit\"at, Max-von-Laue-Str. 1, 60438 Frankfurt am Main, Germany}
\author{J. M. Fern\'andez}
\affiliation{Laboratory of Molecular Fluid Dynamics, Instituto de Estructura de la Materia, CSIC, Serrano 121, 28006, Madrid, Spain}
\author{F. Tramonto}
\affiliation{Laboratorio di Calcolo Parallelo e di Simulazioni di Materia Condensata, Dipartimento di Fisica, Universit\`a degli Studi di Milano, Via Celoria 16, 20133, Milano, Italy}
\author{G. Tejeda}
\author{E. Moreno}
\affiliation{Laboratory of Molecular Fluid Dynamics, Instituto de Estructura de la Materia, CSIC, Serrano 121, 28006, Madrid, Spain}
\author{A. Kalinin}
\affiliation{Institut f\"ur Kernphysik, J. W. Goethe-Universit\"at, Max-von-Laue-Str. 1, 60438 Frankfurt am Main, Germany}
\author{M. Nava}
\affiliation{Laboratorio di Calcolo Parallelo e di Simulazioni di Materia Condensata, Dipartimento di Fisica, Universit\`a degli Studi di Milano, Via Celoria 16, 20133, Milano, Italy}
\affiliation{Computational Science, Department of Chemistry and Applied Biosciences, ETH Zurich, USI Campus, Via Giuseppe Buffi 13, CH-6900 Lugano, Switzerland}
\author{D. E. Galli}
\affiliation{Laboratorio di Calcolo Parallelo e di Simulazioni di Materia Condensata, Dipartimento di Fisica, Universit\`a degli Studi di Milano, Via Celoria 16, 20133, Milano, Italy}
\author{S. Montero}
\affiliation{Laboratory of Molecular Fluid Dynamics, Instituto de Estructura de la Materia, CSIC, Serrano 121, 28006, Madrid, Spain}
\author{R.E. Grisenti}
\affiliation{Institut f\"ur Kernphysik, J. W. Goethe-Universit\"at, Max-von-Laue-Str. 1, 60438 Frankfurt am Main, Germany}
\affiliation{GSI Helmholtzzentrum f\"ur Schwerionenforschung, Planckstr. 1, 64291 Darmstadt, Germany}

\begin{abstract}

By means of Raman spectroscopy of liquid microjets we have investigated the crystallization process of supercooled quantum liquid mixtures composed of parahydrogen (pH$_2$) diluted with small amounts of up to 5\% of either neon or orthodeuterium (oD$_2$), and of oD$_2$ diluted with either Ne or pH$_2$. We show that the introduction of Ne impurities affects the crystallization kinetics in both the pH$_2$-Ne and oD$_2$-Ne mixtures in terms of a significant reduction of the crystal growth rate, similarly to what found in our previous work on supercooled pH$_2$-oD$_2$ liquid mixtures [M. K\"uhnel et {\it al.}, Phys. Rev. B \textbf{89}, 180506(R) (2014)]. Our experimental results, in combination with path-integral simulations of the supercooled liquid mixtures, suggest in particular a correlation between the measured growth rates and the ratio of the effective particle sizes originating from quantum delocalization effects. We further show that the crystalline structure of the mixture is also affected to a large extent by the presence of the Ne impurities, which likely initiate the freezing process through the formation of Ne crystallites.

\end{abstract}

\maketitle

\section{Introduction}

The study of crystallization in supercooled liquids is of prime importance for understanding the fundamental mechanisms of crystal growth. It is relevant to diverse research areas such as microstructural control in alloys engineering or the formation of ice in the atmosphere. In addition, it is of increasing interest in the context of the glass transition\cite{Ediger}, as the maximum rates of crystal growth may themselves provide insights into the glass-forming ability of supercooled liquids \cite{Ediger2,Nascimento,Tang,Orava}. Among the systems that can be used as test beds for our current microscopic understanding of crystallization in supercooled liquids are binary mixtures. The crystallization of binary mixtures differs significantly from that of pure fluids, displaying a rich phase behavior in dependence of particle size ratio and composition. For example, understanding why for some compositions certain metallic binary alloys invariably form crystal phases that compete with the glass formation \cite{Wang} is an open and fundamental question for understanding crystallization and its interplay with the glass-forming ability.

The dynamics and crystallization of supercooled liquid binary mixtures in dependence on composition and particle size disparity have been investigated by classical molecular dynamics simulations of systems interacting through a simple Lennard-Jones (LJ) pair potential \cite{JRFernandez,Coslovich,Valdes,Banerjee}. However, systematic experimental studies of such simple atomic and molecular systems have so far remained out of reach. Here, we present experimental results, obtained by employing the liquid microjet technique in combination with Raman light scattering \cite{Kuehnel}, on the crystallization process of supercooled binary liquid mixtures of either pH$_2$ or oD$_2$ diluted with small amounts of Ne impurities.

The present work was partly motivated by our recent investigations of the crystallization kinetics of supercooled pH$_2$-oD$_2$ liquid mixtures \cite{Kuehnel2}. The most striking observation in those studies was the strong dependence of the crystallization rate on composition: starting with a pure pH$_2$ system, the crystal growth was found to slow down considerably with increasing amount of oD$_2$, reaching a maximum in the crystallization time for the $\left({\rm pH_2}\right)_{80}\left({\rm oD_2}\right)_{20}$ mixture. This behavior may appear surprising because the intermolecular interaction potential is isotope independent \cite{Silvera}. However, quantum delocalization due to zero-point motion increases the radius of the pH$_2$ and oD$_2$ molecules to a different extent due to their different masses, resulting in an effective oD$_2$ to pH$_2$ particle size ratio smaller than unity. This feature led us to interpret the observed growth rates in terms of packing effects linked to the interplay between composition and particle size ratio in these quantum binary mixtures \cite{Kuehnel2}.

Here we have extended the above studies to supercooled pH$_2$-Ne and oD$_2$-Ne liquid mixtures, addressing in particular relevant structural aspects of the crystallization process in the Ne mixtures as well as in pH$_2$-oD$_2$ mixtures not taken into account in our recent work\cite{Kuehnel2}. In principle, a Ne atom can be viewed as an isotopic impurity when mixed to pH$_2$ or oD$_2$ because its interatomic spherical potential is very similar to those of pH$_2$ and oD$_2$. Indeed, the LJ potential parameters (potential well depth $\epsilon$ and interaction length $\sigma$) for the H$_2$--H$_2$ interaction are $\sigma = 2.96$ \AA $ $ and $\epsilon = 34.2$ K, while for H$_2$--Ne they are $\sigma = 2.8745$ \AA $ $ and $\epsilon = 35.49$ K (Ref. \onlinecite{Challa}). However, Ne is 10 times heavier than pH$_2$ and, as a consequence, the magnitude of quantum delocalization is much smaller, resulting in a significantly smaller effective size when compared to pH$_2$ or oD$_2$. The pH$_2$-Ne and oD$_2$-Ne liquid mixtures investigated here represent thus ideal systems to further explore the effect of composition and particle size ratio on the crystallization of supercooled binary liquid mixtures.

\section{Experimental details}

\begin{figure}[t]
\begin{center}
\includegraphics[width=9cm]{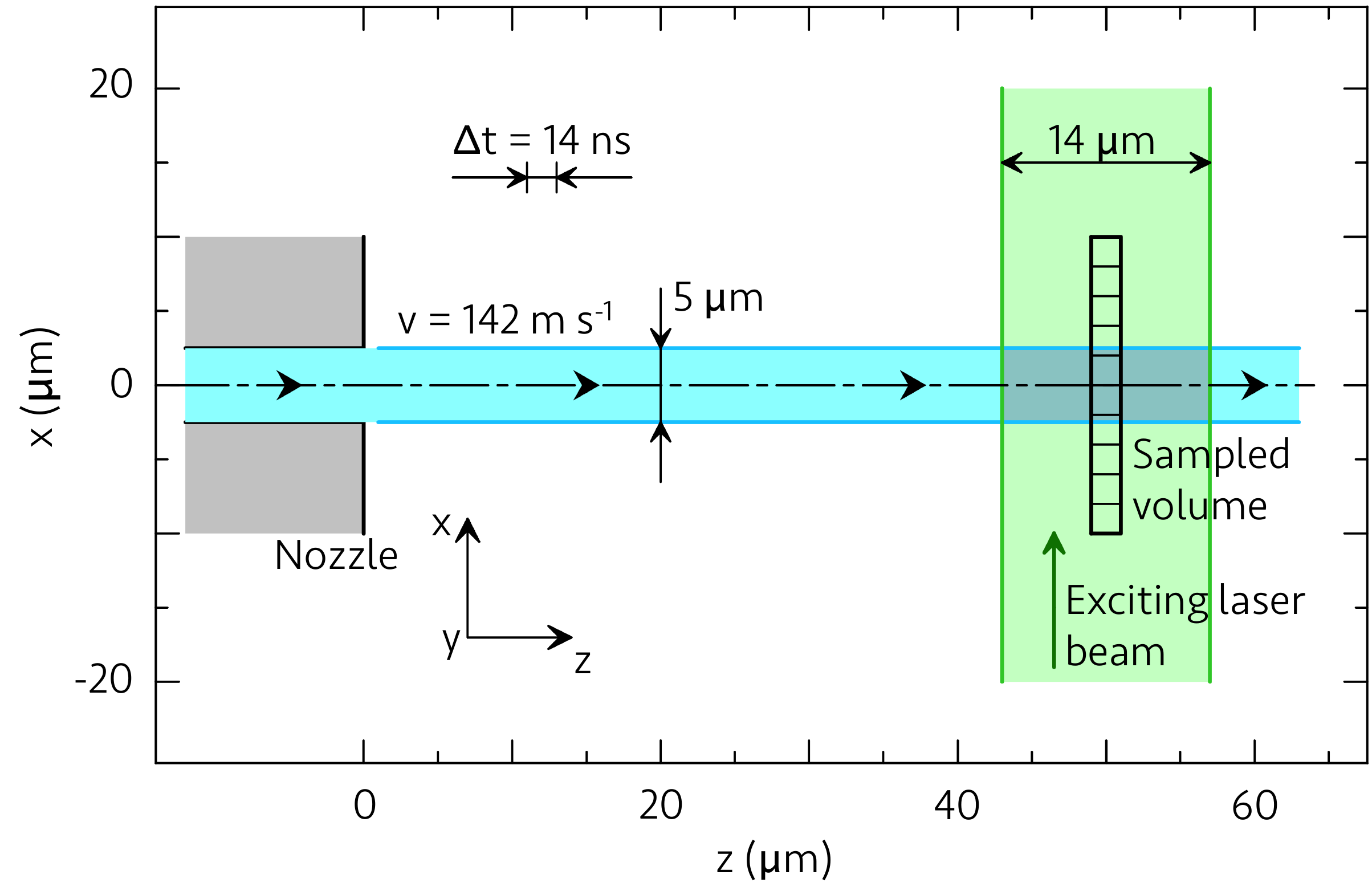}
\caption{\label{exp} Schematic view of the experiment, drawn to scale. The liquid microjet (light blue) flows along the $z$ axis. The exciting laser beam (green) polarized parallel to either the $z$ or $y$ axes propagates along the $x$ direction. Raman scattering is collected along the $y$ direction. The depicted sampled volume is the projection, onto the scattering plane, of the spectrograph entrance slit and of the active area of the CCD detector, showing the space resolution.}
\end{center}
\end{figure}

The supercooled liquid mixtures were produced by the microjet technique and probed by Raman scattering, as described in Refs \onlinecite{Kuehnel,Fernandez}.  A schematic view of the experimental setup is shown in Fig.~\ref{exp}. The liquid jet, which propagates at a speed $v\simeq \sqrt{2 P_0 / \rho}$ ($P_0$ is the source pressure and $\rho$ is the liquid density), rapidly cools well below the melting temperature until it crystallizes, producing a continuous solid filament several cm long \cite{Kuehnel}. A crucial feature of our approach is represented by the correspondence between the distance along the jet propagation direction, $z$, and time, $t=z/v$, allowing for a probe of the crystallization kinetics with sub-microsecond time resolution.

The liquid mixtures were continuously injected into vacuum through a 5 $\mu$m-diameter glass capillary nozzle. The glass capillary, mounted on a microactuators stage allowing a displacement of the entire nozzle assembly, and thus of the probed volume (Fig.~\ref{exp}), along the $x$, $y$, and $z$ directions with an accuracy of better than 1~$\mu$m, was cooled by a continuous flow liquid helium cryostat, and its temperature $T_0$ was actively stabilized within $\pm 0.1$~K. The vacuum chamber was evacuated by a 2000 l/s turbo molecular pump providing a background pressure below $3 \times 10^{-3}$~mbar. The nuclear spin variants pH$_2$ and oD$_2$ were produced by continuous catalytic conversion at 17 and 22~K, respectively, from 99.9999\% and 99.9\% purity natural H$_2$ and D$_2$ gases, respectively, resulting in 99.8\% and 97.5\% purity pH$_2$ and oD$_2$, respectively, the rest being represented by odd-$J$ molecules. The room temperature gas streams of pH$_2$, oD$_2$, and 99.998\% purity Ne were then mixed at the specific ratios by two mass flow controllers, one for each component, working at a minimum flow rate of 20 normal-ml/min. The equilibrium solubility of neon in hydrogen in the liquid phase is limited to about 5\%, and to about twice that value for neon in deuterium, whereas a phase separation occurs at higher Ne content\cite{Brouwer,Barylnik,Galtsov,Belan}. The experimental conditions for all the mixtures investigated here are reported in Table~\ref{table}.

\begin{table}[t]
\caption{\label{table} Relevant experimental parameters (see the text for details) and impurity content for all the systems studied in the present work.}
\begin{center}
\begin{tabular}{cccccc} \hline\hline
 System & Impurity & (mol \%) & $P_0$ (bar) & $T_0$ (K) & $v$ (m/s) \\  \hline
pH$_2$         &  --   &   0   &   8.5       &  16.0     &  142      \\  \hline
oD$_2$         &  --   &   0   &  13.0       &  19.6     &  115      \\  \hline
                      &   Ne   & 0.3 &  9.4       &  17.0     &  147      \\
 pH$_2$-Ne   &   Ne   & 1 &  11.2       &  17.5     &  153      \\
                      &    Ne  & 2 &  10.8       &  18.0     &  137      \\   \hline
                      &   Ne   & 1 &  16.0       &  20.0     &  125      \\
oD$_2$-Ne    &   Ne   & 2 &  17.0       &  20.0     &  123      \\
                      &   Ne   & 5 &  11.8       &  20.0     &   95      \\ \hline
                      & oD$_2$  & 1 &  11.5       &  16.0     &  164      \\
                      & oD$_2$ & 3.2  &   8.9       &  16.0     &  143      \\
 pH$_2$-oD$_2$  & oD$_2$ & 4.6  &   9.3       &  16.0     &  144      \\
                      & pH$_2$ & 2.4  &  16.3       &  20.0     &  130      \\
                     & pH$_2$ &  4.3   &  15.3       &  20.0     &  127      \\ \hline\hline
\end{tabular}
\end{center}
\end{table}
The liquid microjets were probed by recording Raman spectra of the vibrational $Q_1(0)$ and rotational $S_0(0)$ transitions of pH$_2$ and oD$_2$ as a function of $z$. The Ne impurities could not be detected as they are Raman scattering inactive. Raman scattering was excited by a 4~W plane-polarized cw Ar$^+$ laser beam at $\lambda= 514.5$~nm, with an intracavity etalon for single-mode operation, and focused down to a diameter of $\approx 14$~$\mu$m onto the filament. The polarization plane of the exciting laser beam could be rotated by means of a $\lambda/2$ plate at the exit of the laser cavity. During the measurements of the rotational spectra the exciting laser beam was polarized along the $y$ axis (Fig.~\ref{exp}) in order to minimize the stray light resulting from the elastic scattering, whereas for the vibrational spectra it was polarized along the $z$ axis to record the Raman scattering in full.

The Raman signal, which is proportional to the molecular number density, was collected at $90^{\circ}$ with respect to both the laser beam and filament axis, and was focused onto the 20~$\mu$m entrance slit of the spectrometer by an optical system with $\times 9$ magnification. A supernotch filter was used to block part of the very intense Rayleigh scattering from the liquid filament. The spectrometer, with one meter focal length, was equipped with a 2360 groove/mm grating and a back-illuminated CCD detector with $20 \times 20$~$\mu$m$^2$ pixels cooled by liquid nitrogen to 153 K. Its spectral resolution is $\approx 0.11$~cm$^{-1}$ at 4150~cm$^{-1}$ (pH$_2$ vibrational band) and $\approx 0.24$~cm$^{-1}$ at 180~cm$^{-1}$ (oD$_2$ rotational band). As illustrated in Fig.~\ref{exp}, the space resolution along the jet axis was determined by the $\approx 2$~$\mu$m projection of the spectrograph slit onto the microjet. Along the $x$ radial direction 10 stripes, each 2 pixels high, were read onto the CCD detector, allowing for a radial sampling of the jet. The Raman spectra presented here were obtained by adding up just over the illuminated stripes.

\section{Results}

\subsection{Crystallization kinetics}

\begin{figure}[t]
\begin{center}
\includegraphics[width=10cm]{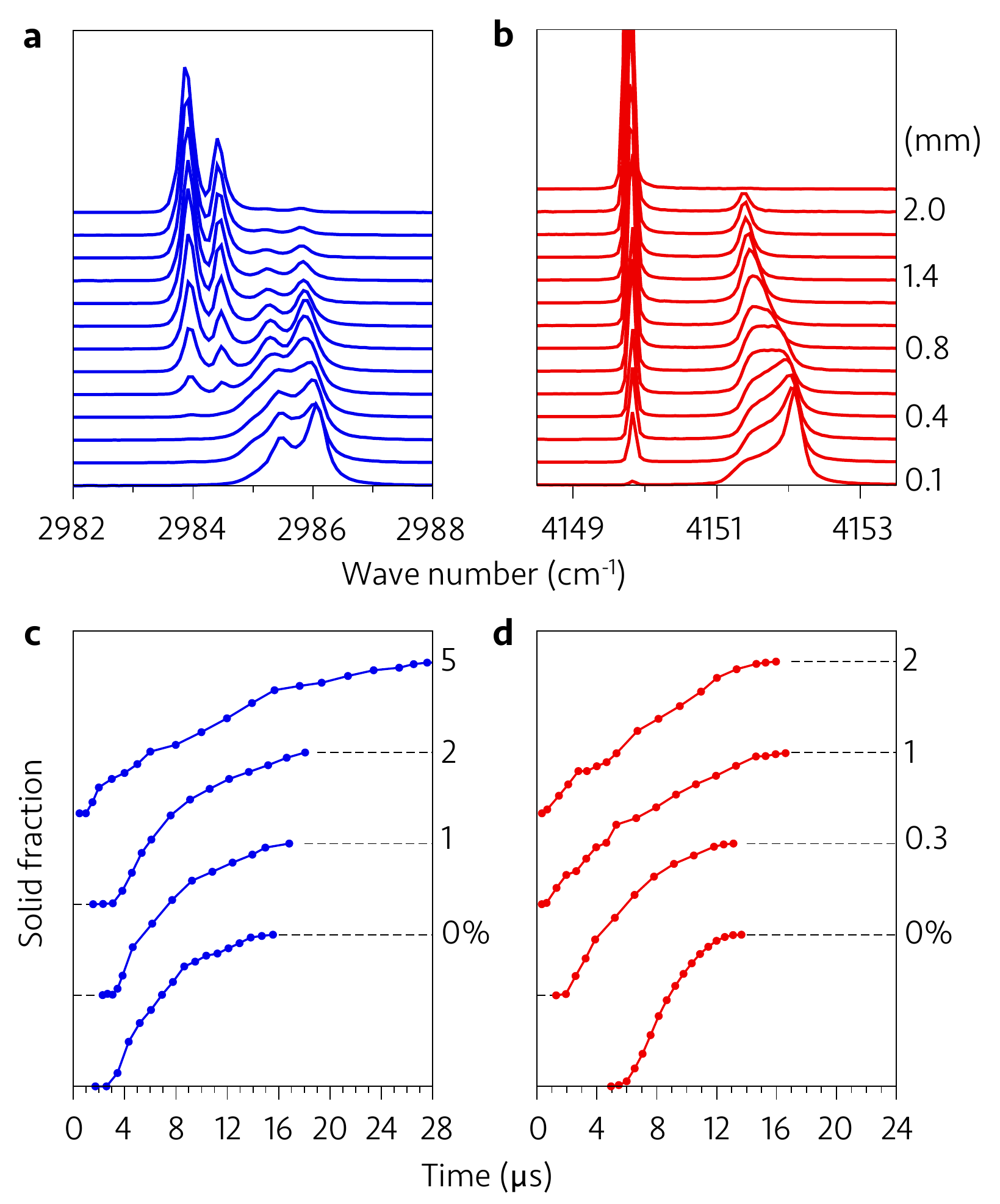}
\caption{\label{vib}(a), (b) Selected normalized vibrational Raman spectra of oD$_2$ (a) and pH$_2$ (b) measured as a function of the distance $z$ from the orifice (right scale) for the 1\% Ne mixtures. The double-line shape of the oD$_2$ bands visible in (a) results from the presence of less than 3\% of $J=1$ pD$_2$ molecules, with a 50-fold enhancement in the Raman scattering intensity with respect to the $J=0$ molecules\cite{Kozioziemsky}. Panels (c) and (d) show the time evolution of the solid fractions extracted from the vibrational bands for oD$_2$ and pH$_2$, respectively, in mixtures with mole percentages indicated on the right axis. The solid fractions range from 0 to 1, as indicated by the dashed lines on the left and on the right of the experimental curves, respectively.}
\end{center}
\end{figure}

Examples of vibrational spectra are shown in Figs~\ref{vib}(a) and \ref{vib}(b) for the $\left({\rm oD_2}\right)_{99}\left({\rm Ne}\right)_{1}$ and $\left({\rm pH_2}\right)_{99}\left({\rm Ne}\right)_{1}$ mixtures, respectively, evidencing the phase transition from the liquid to the solid. As mentioned above, the crystallization behavior of the Ne impurities could not be addressed during the present work because they are not Raman active. The vibrational spectra allow extracting the solidified fraction of the sampled filament volume. The evolution of the oD$_2$ and pH$_2$ solid fractions measured for all the oD$_2$-Ne and pH$_2$-Ne mixtures investigated here are shown in Figs \ref{vib}(c) and \ref{vib}(d), respectively, where the axial distance has been converted into time as explained above.

Overall, the crystallization kinetics features of Fig.~\ref{vib} are qualitatively similar to those reported recently for pH$_2$-oD$_2$ mixtures \cite{Kuehnel2}, yet the crystallization slowdown observed upon addition of small amounts of Ne impurities is strongly enhanced. In Fig. \ref{vib}(d) we see that the presence of merely 0.3\% Ne results in a significantly earlier start of the crystallization compared to pure pH$_2$. In the case of the $\left({\rm pH_2}\right)_{99}\left({\rm Ne}\right)_{1}$ and $\left({\rm pH_2}\right)_{98}\left({\rm Ne}\right)_{2}$ mixtures the vibrational peak corresponding to solid pH$_2$ appears already within the first 100 $\mu$m distance from the orifice [Fig. \ref{vib}(b)], whereas the solidification process lasts about twice with respect to the pure pH$_2$ case. In the case of the oD$_2$-Ne mixtures [Fig. \ref{vib}(c)] the presence of up to 2\% Ne does not actually lead to a clear earlier onset of freezing with respect to the pure case, though the subsequent crystal growth slows down with increasing Ne concentration in a similar fashion as for the pH$_2$-Ne mixtures. We recall that an earlier onset of freezing has been also observed in the pH$_2$-oD$_2$ mixtures \cite{Kuehnel2}, especially upon addition of oD$_2$ impurities.

\subsection{Crystal structure}

The vibrational Raman spectra provide insights into the crystallization kinetics but they do not offer any straightforward information on the structure of the growing crystal. Such information can be retrieved from the rotational spectra, which we have probed for the pH$_2$-Ne and oD$_2$-Ne mixtures, as well as for diluted pH$_2$-oD$_2$ mixtures. Since the rotational spectrum of the liquid is a broad band in all cases, we focus here on the structural properties after completion of the freezing process. The evolution of the rotational excitations during the liquid-to-solid phase transition in supercooled pH$_2$ has been discussed elsewhere \cite{Kuehnel}. 

\begin{figure}[t]
\begin{centering}
\includegraphics[width=9cm]{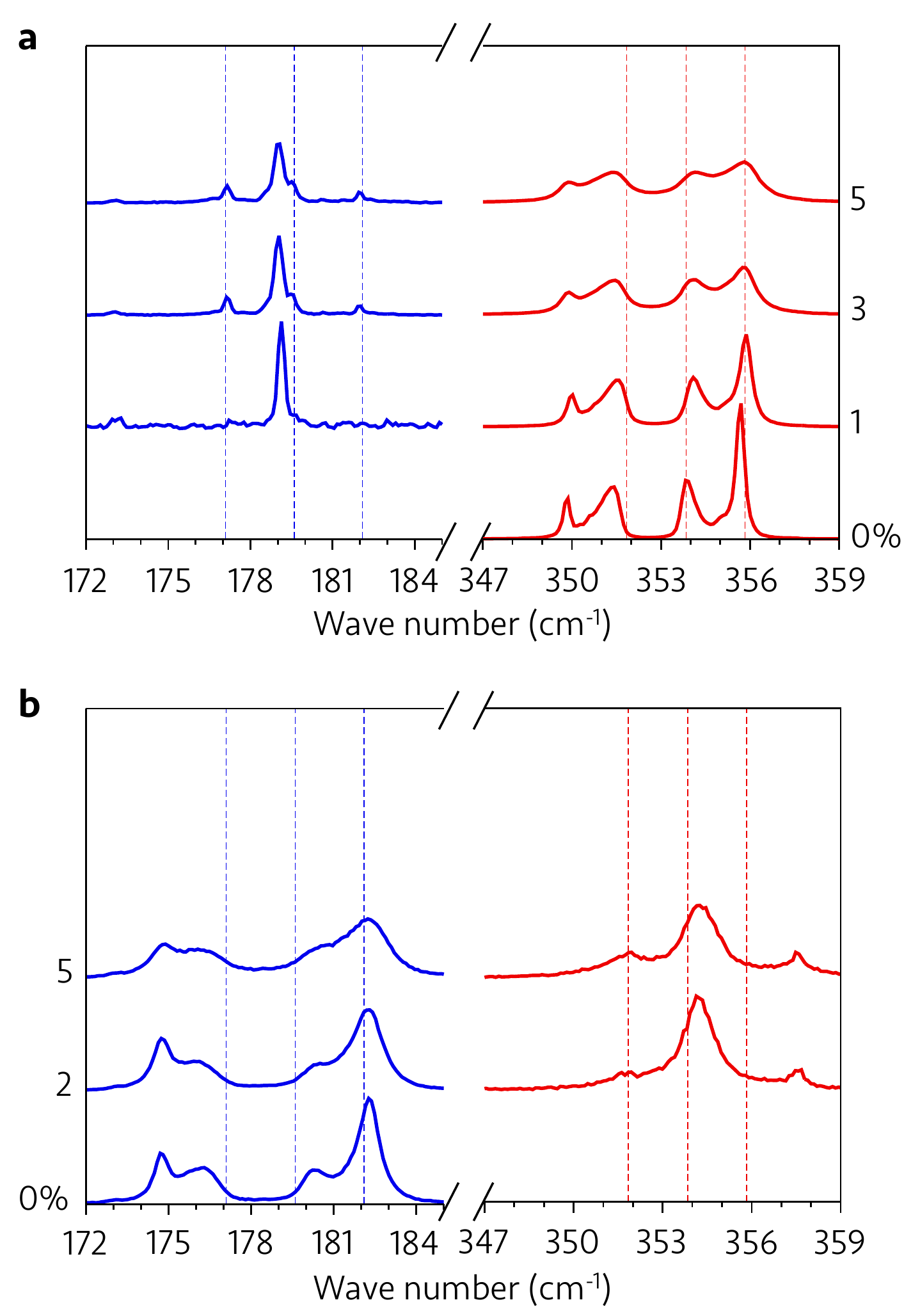}
\caption{\label{rot}Normalized rotational Raman spectra of oD$_2$ (blue) and pH$_2$ (red) in completely solidified filaments of diluted pH$_2$-oD$_2$ mixtures. The impurity component is oD$_2$ and pH$_2$ in (a) and (b), respectively. The oD$_2$ and pH$_2$ impurity percentages are indicated on the right and left scales in (a) and (b), respectively. Vertical red and blue dashed lines indicate the position of the Raman lines measured for equilibrium hcp crystals of pure pH$_2$ and oD$_2$, respectively. The small peaks visible around 173 cm$^{-1}$ in (a) and around 358 cm$^{-1}$ in (b) are plasma lines from the Ar$^+$ laser.}
\end{centering}
\end{figure}
In Fig.~\ref{rot}(a) and \ref{rot}(b) we present the rotational spectra measured for the pH$_2$-oD$_2$ mixtures with oD$_2$ and pH$_2$ as impurity, respectively; the spectra were recorded at a distance of about 4 mm downstream from the orifice, corresponding to a propagation time of 40 $\mu$s, long enough for a complete crystallization in all cases (see Fig.~\ref{vib}). The rotational bands of both pure solid pH$_2$ [Fig.~\ref{rot}(a)] and oD$_2$ [Fig.~\ref{rot}(b)] filaments are characterized by four distinct peaks. This feature has been attributed to a random hexagonal closed packed (rhcp) structure, i.e., an alternating stacking of hexagonal closed-packed (hcp) and face-centered cubic (fcc) crystal domains \cite{Kuehnel}. In fact, bulk solid pH$_2$ and oD$_2$ slowly grown from the melt always exhibit an hcp structure, whose rotational Raman spectrum consists of three lines [the red and blue dashed lines in Fig.~\ref{rot}(a) and \ref{rot}(b), respectively]. However, while the hcp lattice is the most stable crystal structure, it was experimentally found that bulk solid pH$_2$ and oD$_2$ grown fast from the vapor at very low temperatures also form the fcc crystal lattice, which differs in energy from the hcp lattice only by a factor of $\sim 10^{-5}$ (Ref. \onlinecite{Silvera}). The rotational Raman band of the pH$_2$ fcc structure consists of two peaks at $\approx 350$ cm$^{-1}$ and $\approx 356$ cm$^{-1}$, the latter overlapping with one peak corresponding to the hcp structure. The addition of small amounts of the impurity leads to a broadening of the rotational features, as shown in Fig.~\ref{rot}(a) and \ref{rot}(b), while maintaining the coexistence of the hcp and fcc lattices. This broadening of the rotational lines upon dilution with either oD$_2$ or pH$_2$ has been observed and interpreted previously as the result of the coupling between the rotational states in the solid mixtures \cite{Kozioziemsky}.

Turning now to the analysis of the rotational excitations relative to the impurity some striking features appear. For instance, we see in Fig.~\ref{rot}(a) that at the lowest concentration of 1\% oD$_2$ the oD$_2$ rotational spectrum exhibits a single line at $\approx 179$ cm$^{-1}$. This line is very sharp and its wavenumber and width match those of the gas phase. This suggests that most of the oD$_2$ impurities occupy sites with no nearest neighbor oD$_2$ molecules, being able to rotate freely as a results of the larger average distance between the pH$_2$ molecules in the crystal matrix. With increasing oD$_2$ impurity concentration up to about 5\%, we observe the gradual appearance of three additional sharp lines at the wavenumbers of the oD$_2$ bulk hcp crystal lattice \cite{Silvera}. This can be rationalized in terms of nearest neighbor $\left({\rm oD_2}\right)_2$ pairs surrounded by pH$_2$ molecules, where the magnetic degeneracy of the rotational states is split along the intermolecular axis, giving rise to the hcp triplet \cite{Kranendonk}.

\begin{figure}[t]
\begin{center}
\includegraphics[width=10cm]{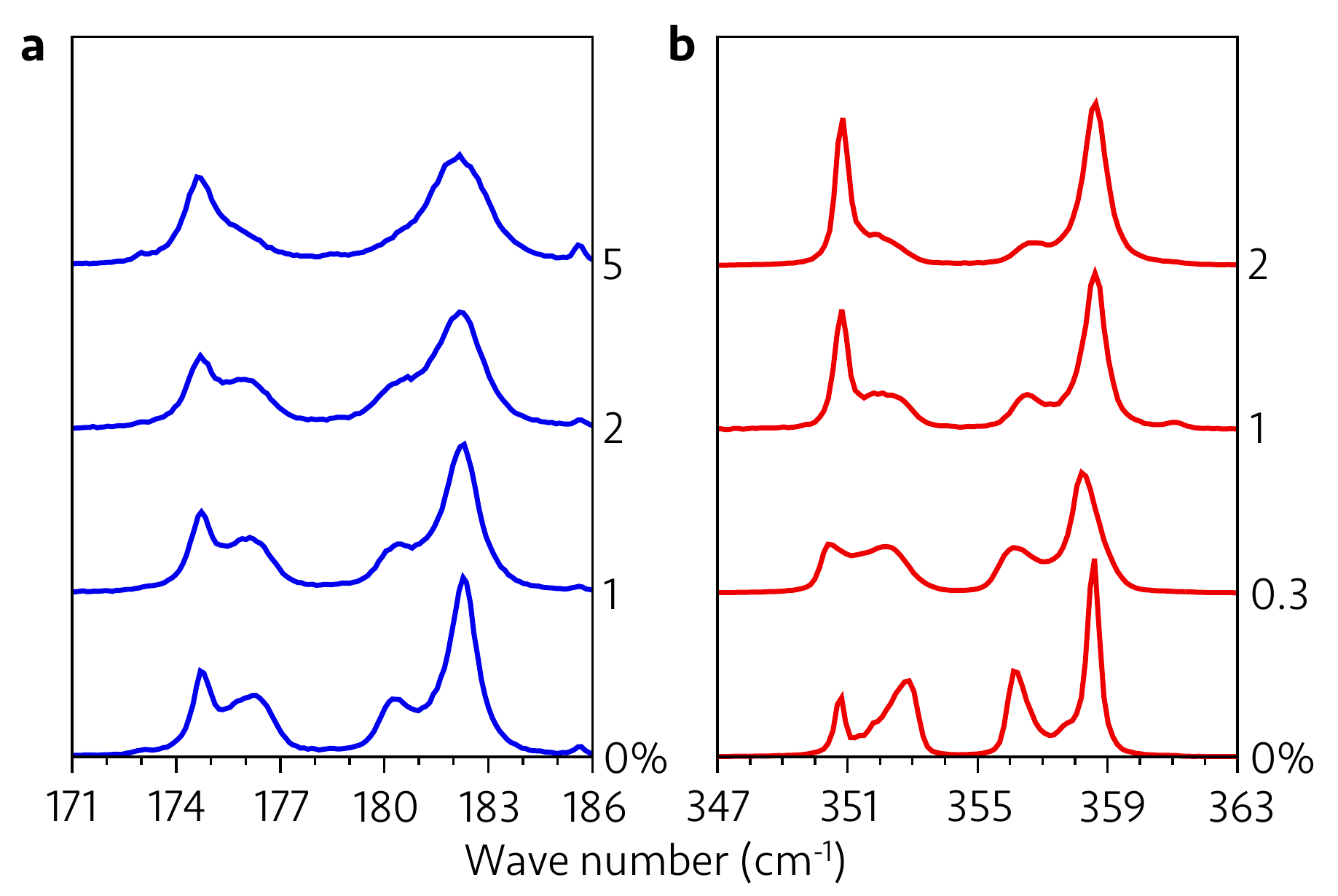}
\caption{\label{rot2}Normalized rotational Raman spectra of oD$_2$ (a) and pH$_2$ (b) in completely solidified filaments of diluted pH$_2$-Ne and oD$_2$-Ne mixtures, respectively. The Ne impurity content is indicated on the right scale.}
\end{center}
\end{figure}
In the case of pH$_2$-oD$_2$ mixtures with pH$_2$ as impurity [Fig.~\ref{rot}(b)] we find a similar trend, yet with noticeable differences. The pH$_2$ rotational spectrum at the lowest concentration of 2\% pH$_2$  is also characterized by a main peak whose wavenumber agrees with that of the free molecule, indicating that most of the pH$_2$ molecules are surrounded by oD$_2$ nearest neighbors. However, this Lorentz-shaped peak is significantly broader than in the case of isolated oD$_2$ impurities [Fig.~\ref{rot}(a)] because the broader zero-point wave function of a pH$_2$ molecule explore part of the repulsive potential of surrounding oD$_2$ molecules, affecting its rotational motion and thus giving rise to a broader Raman peak. The small side peak at $\approx 352$ cm$^{-1}$, which grows with increasing pH$_2$ concentration, matches the lowest-energy excitation line of the hcp triplet in the bulk (red dashed lines). This suggests again that a fraction of the pH$_2$ molecules in the lattice is arranged in nearest neighbor $\left({\rm pH_2}\right)_2$ pairs, the remaining two hcp peaks being probably hidden by the broad central peak. 

The rotational spectra of the solidified filaments containing Ne as impurity are presented in Fig.~\ref{rot2}, evidencing remarkable differences with respect to the case of the pH$_2$-oD$_2$ mixtures. In Fig.~\ref{rot2}(b) we can see that the intensities of the hcp lines of pH$_2$ at $\simeq 353$ and $\simeq 356$ cm$^{-1}$, for a Ne concentration as low as 0.3\%, drop to approximately half of their value for pure pH$_2$, and become vanishingly small for the 2\% Ne mixture. Simultaneously, the intensity of the fcc line at 351 cm$^{-1}$ increases, becoming the dominant feature, together with the line at 359 cm$^{-1}$, at the highest Ne concentration of 2\%. A similar trend is observed for the oD$_2$-Ne mixtures [Fig.~\ref{rot2}(a)], though the two dominant peaks at $\simeq 175$ and $\simeq 182$ cm$^{-1}$ in the oD$_2$ rotational spectra at the highest Ne concentrations are broader than those in the pH$_2$ spectra of Fig.~\ref{rot2}(b). Overall, the data of Fig.~\ref{rot2} indicate that by adding small amounts of Ne to either pH$_2$ or oD$_2$ the crystal structure changes from an rhcp configuration to a dominant fcc structure, which is the equilibrium configuration of solid neon. We point out that the slow crystallization from equilibrium H$_2$-Ne and oD$_2$-Ne liquid mixtures results in a physical separation into pure Ne and H$_2$ (D$_2$) solid phases, where the amount of the fcc domains is negligibly small at such low Ne concentrations \cite{Barylnik,Galtsov,Belan}.

\section{Discussion}

\begin{figure}[t]
\begin{centering}
\includegraphics[width=9cm]{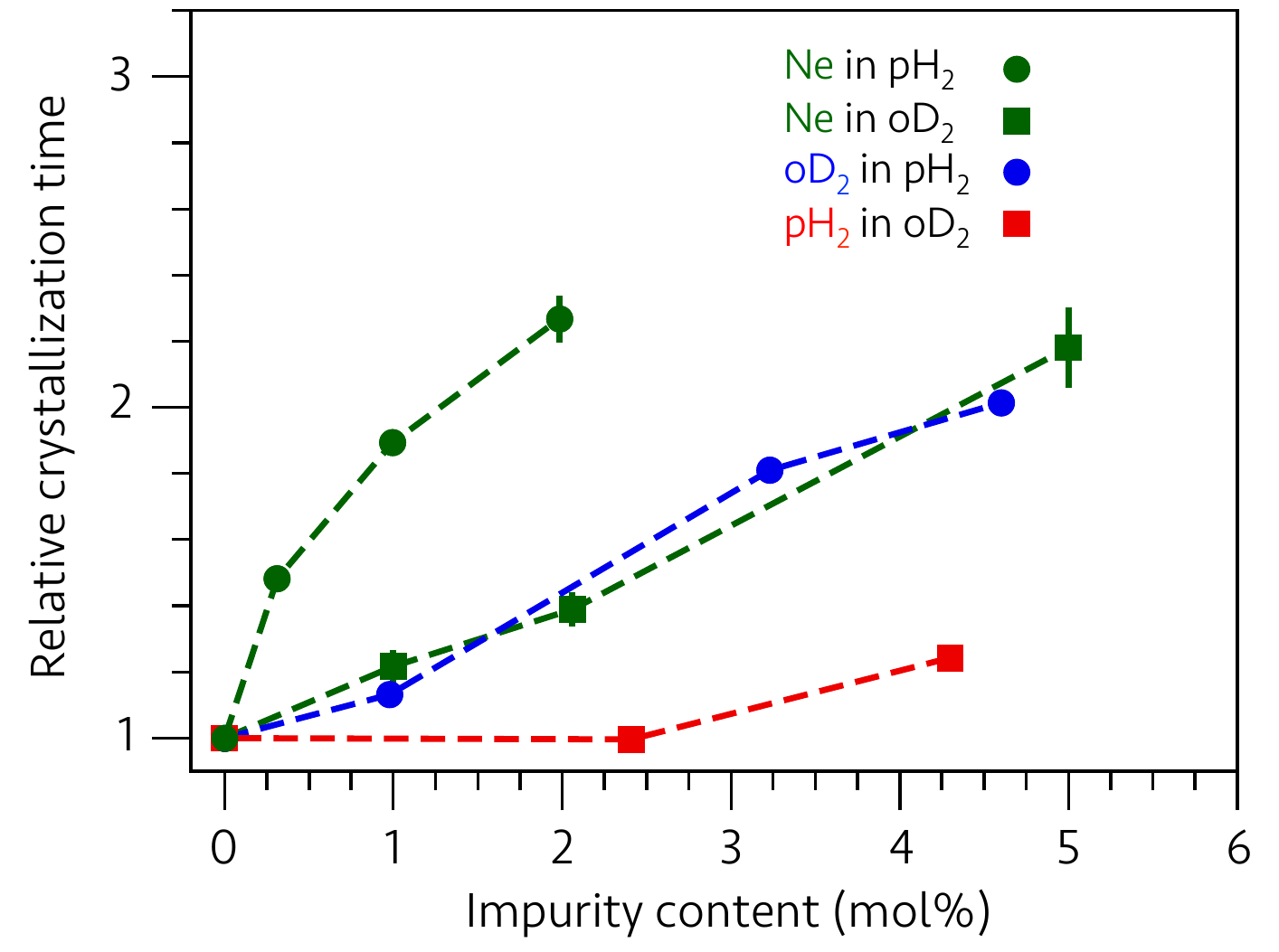}
\caption{\label{summary} Crystallization time for microjets of diluted pH$_2$-Ne and oD$_2$-Ne liquid mixtures as determined from the pH$_2$ (green circles) and oD$_2$ (green squares) solid fraction curves of Figs \ref{vib}(c) and \ref{vib}(d), respectively. Also shown are our previous data obtained for pH$_2$-oD$_2$ mixtures \cite{Kuehnel2} with oD$_2$ (blue circles) and pH$_2$ (red squares) as the impurity species. The plotted data indicate relative values with respect to the crystallization time measured for either pure pH$_2$ or pure oD$_2$ liquid jets. The dashed lines are guides to the eye.}
\end{centering}
\end{figure}

At the microscopic scale the process of solidification from the melt is governed by crystal nucleation and the subsequent crystal growth. The overall picture that emerges from our crystallization kinetics data is that for the pure pH$_2$ and oD$_2$ systems the limiting factor for crystallization is the formation of a critical crystallite; once a critical crystallite has formed, the subsequent crystal growth is fast (the growth rate for pure pH$_2$ is about 30 cm$\,$s$^{-1}$). However, by adding small amounts of Ne atoms the nucleation time becomes shorter and the limiting factor for crystallization in this case is the reduced crystal growth rate. We present in Fig.~\ref{summary} a summary of the crystallization kinetics results by plotting the total crystallization time of the solvent, relative to the pure case, in dependence of the impurity content for the pH$_2$-Ne and oD$_2$-Ne mixtures, as well for the pH$_2$-oD$_2$ mixtures\cite{Kuehnel2}. One can clearly distinguish two limiting cases, represented, on one side, by the barely significant increase of the crystallization rate of oD$_2$ upon dilution with pH$_2$ (red squares) and, on the other side, by the strong dependence on the impurity content of the crystallization rate of pH$_2$ upon dilution with Ne mixtures (green circles).

According to the classical kinetic theory \cite{Jackson} the crystal growth rate is expressed in terms of thermodynamic and kinetic factors as
\begin{equation}
u(T) = k(T)\left[1-e^{-\Delta G(T)/k_BT}\right]
\end{equation}
where $T$ is the temperature, $\Delta G(T)$ is is the difference in Gibbs free energy (per molecule) of the liquid and the crystal and is therefore the driving force for crystallization, $k(T)$ is the crystal deposition rate at the liquid/crystal interface, and $k_B$ is Boltzmann's constant. For the isotopic pH$_2$-oD$_2$ mixtures $\Delta G(T)$ can be computed from the experimental heat capacity data for the pure pH$_2$ and oD$_2$ systems, and just a slight dependence on composition was found \cite{Kuehnel2}. Due to the lack of experimental heat capacity data a similar direct determination of $\Delta G(T)$ was not possible for mixtures containing Ne. However, due to their nearly isotopic nature we do not expect that the thermodynamic factor plays a major role when describing the crystallization kinetics of pH$_2$-Ne and oD$_2$-Ne mixtures. Accordingly, the observed composition dependence of the crystal growth rate must be contained in the kinetic term $k(T)$.

The empirical Wilson-Frenkel model assumes that crystal growth is an activated process in which mass transport, expressed in $k(T)$ through the diffusion constant, represents the most important limiting factor for the growth rate \cite{Jackson}. But how does mixing affect the transport properties, and consequently the crystallization kinetics, to the extent reported here? A number of simulation studies proposed that a lower particle diffusivity can be associated to dense packing effects resulting from the development in the liquid of some local coordination geometries \cite{Royall}. Recently, the composition dependence of transport properties observed in experiments with binary metallic liquids has been explicitly linked to corresponding changes in the density of packing \cite{Stueber,Kuhn}. Indeed, from a pure geometrical point of view the packing efficiency in a mixture of hard spheres of specific size ratio is a function of composition \cite{Hopkins}. In this respect, path Integral Monte Carlo (PIMC) simulations of supercooled pH$_2$-oD$_2$ liquid mixtures have evidenced that, as a result of the different effective sizes of the pH$_2$ and oD$_2$ isotopes, there is a slightly higher probability to find icosahedral-like order around an oD$_2$ molecule than around a pH$_2$ molecule \cite{Kuehnel2}. Among the possible local geometrical structures icosahedra are those that are most densely packed and have been associated to the glass-forming ability of supercooled liquids \cite{Hirata}.

\begin{figure}[t]
\begin{center}
\includegraphics[width=8cm]{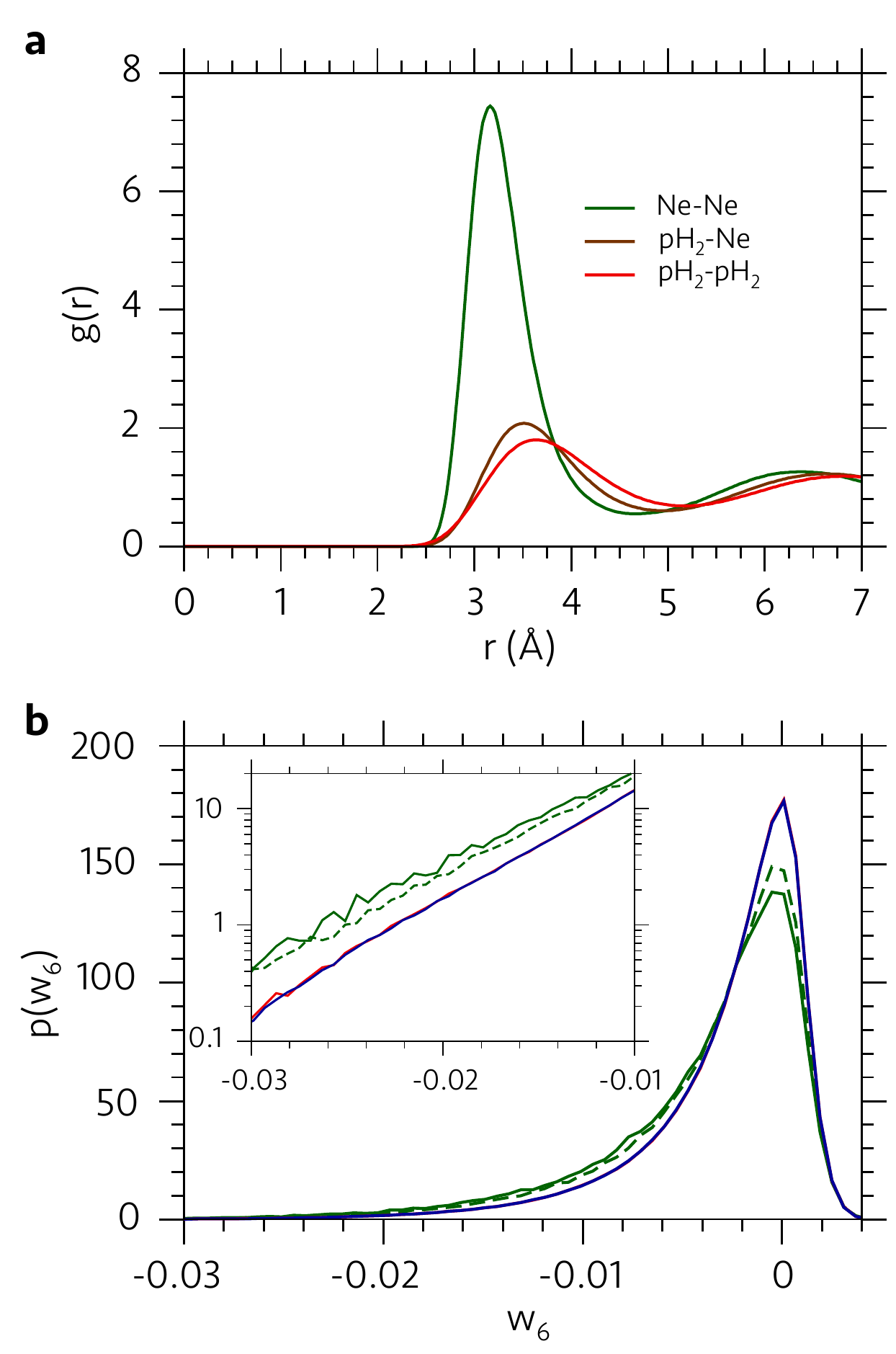}
\caption{\label{sim}PIMC simulations results for pH$_2$-Ne and oD$_2$-Ne with 3\% Ne computed at $T = 13$ K and $T = 17$ K, respectively. (a) Partial radial distribution functions $g(r)$ representing the three pair correlations for the pH$_2$-Ne mixture. (b) Probability distribution $p(w_6)$ for icosahedral-like order for the pH$_2$ (red), oD$_2$ (blue), and Ne (green) particles. The green solid line refers to the pH$_2$-Ne mixture whereas the dashed lines refers to the oD$_2$-Ne mixture. The inset shows an enlarged view of the tail on a logarithmic scale, emphasizing the differences between the different species species at high negative values of the local bond order parameter $w_6$. Note the similar probabilities for local icosahedral packing around a pH$_2$ or oD$_2$ molecule.}
\end{center}
\end{figure}
We have performed analogous PIMC simulations of the Ne mixtures (see Appendix for details), and the main results are shown in Fig.~\ref{sim}. In Fig.~\ref{sim}(a) we plot the partial radial pair distribution functions computed for a $\left({\rm pH_2}\right)_{97}\left({\rm Ne}\right)_3$ mixture. A local bond order analysis (see Appendix) shows that the observed differences in the average static inter-particle correlations translate into a higher degree of icosahedral-like order around a Ne atom than around a pH$_2$ molecule, as shown in Fig.~\ref{sim}(b). However, as a result of the higher Ne mass compared to oD$_2$ this tendency for local ordering around a Ne impurity in the pH$_2$-Ne mixtures is more pronounced than that found around an oD$_2$ molecule in pH$_2$-oD$_2$ mixtures. Furthermore, the comparison in Fig.~\ref{sim}(b) of the distributions computed for the pH$_2$-Ne and oD$_2$-Ne mixtures shows also that the probability for local ordering around a Ne impurity in the former mixture is appreciably higher than in the latter mixture. These results thus further support a correlation between dense packing and slower crystallization kinetics, offering in addition a suggestive trend: the magnitude of the observed reduction in the crystal growth rate upon mixing appears to be roughly correlated with the effective particle size ratio for the specific mixture. In fact, one can obtain an indirect estimation of the effective particle size ratio by comparing the first peak positions in the partial radial distribution functions computed along the PIMC simulations; it turns out that the ratio of the first peak positions in the pH$_2$-Ne and pH$_2$-pH$_2$ radial distributions is about 0.966, to which corresponds the strongest crystallization slow down (Fig.~\ref{summary}). This ratio becomes about 0.971 for the oD$_2$-Ne system, whereas for the pH$_2$-oD$_2$ mixture\cite{Kuehnel2} the ratio of the first peak positions is about 0.993 and 1.014 with oD$_2$ and pH$_2$ as impurity, respectively.

The rotational Raman spectra reported here offer additional insights into the crystallization process in supercooled quantum binary mixtures. We have seen (Fig.~\ref{rot2}) that the presence of small amounts of Ne impurities tends to force both pH$_2$ and oD$_2$ to assume a crystal structure that can be associated to an fcc lattice, which is the equilibrium configuration of solid Ne. It is unlikely that the presence of a tiny amount of Ne impurities at the 1\% level and randomly distributed in the lattice would be able to modify the host pH$_2$ or oD$_2$ crystal structure to such an extent. The immediate interpretation of these observations is that the Ne impurities trigger the crystallization process with the formation of Ne-rich fcc crystallites that act as nucleation sites and thus subsequently grow into the surrounding pH$_2$- or oD$_2$-rich supercooled liquid by keeping their initial crystal structure. The growth into an fcc crystalline structure can be facilitated by the small energy difference between the fcc and hcp lattices. The formation of Ne-rich agglomerates is likely driven by compositional fluctuations combined to the deep supercooling and the strong Ne-Ne correlation [Fig.~\ref{sim}(a)]. That a clear tendency to assume a specific crystal structure is not seen in the pH$_2$-oD$_2$ mixtures (Fig.~\ref{rot}) is due to the fact that both pH$_2$ and oD$_2$ crystallize into an hcp lattice \cite{Silvera}. This simple picture would also explain the earlier start of crystallization upon addition of the Ne impurities, especially evident for the pH$_2$-Ne mixtures [Fig. \ref{vib}(d], as the probability for the formation of large impurity clusters increases with increasing Ne amount. The slight differences in the occurrence of the onset of freezing in the pH$_2$-Ne and oD$_2$-Ne mixtures [compare Figs \ref{vib}(c) and \ref{vib}(d)] may be partly related to the equilibrium D$_2$-Ne phase diagram\cite{Belan}, which shows a slightly decrease of the liquidus temperature at small ($<3$\%) Ne concentrations, in stark contrast to its increase observed in equilibrium H$_2$-Ne mixtures\cite{Brouwer,Barylnik}. As a final comment we note that a possible competition between the hcp and fcc structures, as suggested by our experimental data, might contribute to the slowdown of crystal growth in the pH$_2$-Ne and oD$_2$-Ne mixtures; the frustration of crystallization due to the conflict between different crystalline structures is an aspect that has been discussed in numerical studies of crystallization of binary mixtures \cite{Valdes,Banerjee}

To conclude, we have shown that the combination of liquid microjets and Raman light scattering can provide highly valuable information on the crystallization process in supercooled quantum liquid mixtures. Our results hint in particular at the importance of effects such atomic-level packing efficiency, related to effective particle size disparity originating from quantum delocalization, and crystal structure competition. It would be highly desirable that the present results could be complemented by exploiting the capabilities offered by the liquid microjet technique in combination with state-of-the-art X-ray sources. Such studies would have the potential to shed further light on the physical origin of the findings reported here, and thus on the basic mechanisms of crystal growth.

\section*{Acknowledgments}

We acknowledge financial support by the Deutsche Forschungsgemeinschaft, through Grant No. 593962, and the Spanish Ministerio de Ciencia e Innovaci\'on, through Grants No. FIS2010-22064-C02 and FIS2013-48275-C2. We acknowledge CINECA and the Regione Lombardia award, under the LISA initiative, for the availability of high performance computing resources and support.

\appendix*
\section{Simulation details}

We have simulated pH$_2$-Ne and oD$_2$-Ne mixtures with 3\% Ne by assuming a simple LJ pair potential to model the interaction among the Ne atoms\cite{Challa}, the standard Silvera-Goldman potential\cite{Silvera2} to model the interaction among hydrogen molecules, and the pair potential reported in Ref. \onlinecite{Challa} to model the pH$_2$--Ne  and oD$_2$--Ne interactions. We have simulated systems with $N=300$ particles in boxes with periodic boundary conditions at a density given by the linear relation $\rho = 0.97 \rho_{\mathrm{pH_2},\mathrm{oD_2}} + 0.03 \rho_{{\rm Ne}}$, where $\rho_{\mathrm{pH_2}} = 0.0232$ \AA$^{-3}$, $\rho_{\mathrm{oD_2}} = 0.0263$ \AA$^{-3}$, and $\rho_{\mathrm{Ne}}=0.0373$ \AA$^{-3}$ are the densities close to triple-point conditions of the pure substances.

Within PIMC the quantum system of $N$ particles is mapped onto a classical system of $N$ ring-polymers\cite{Ceperley95}. Each ring-polymer is composed of beads, whose number is fixed by expressing the full density matrix $\rho(R,R^{\prime},T)=\langle R | \exp(-\hat{H}/k_B T)|R'\rangle$ ($\hat{H}$ is the Hamilton operator, $R$ and $R^{\prime}$ are many-body coordinates representing the positions of the $N$ particles, and $T$ is the temperature), via a Trotter decomposition, as a convolution of higher temperature density matrices $\rho(R,R^{\prime},T')$. In our simulations we adopted the pair-Suzuki approximation\cite{Rossi09} for $\rho(R,R^{\prime},T^{\prime})$ with $T^{\prime}=720$ K and assumed a total number of 54 and 41 convolutions for the pH$_2$-Ne and oD$_2$-Ne systems, respectively. The corresponding temperatures are $T=13.09$ K and $T=17.14$ K, respectively, which represent the estimated experimental average filament temperatures of the pure systems\cite{Kuehnel2}; given the low Ne content, the same temperatures have been assumed also for the simulations of the mixtures.

To determine the geometry of the local environment (either crystalline or liquid) around a particular particle of the simulated quantum binary mixtures we used the identification schema based on the local bond order (LBO) parameters introduced by Steinhardt et al.\cite{Steinhardt83}, modified by averaging the LBO parameters over the nearest neighbors of the analyzed particle and the particle itself \cite{Lechner08}. Since quantum delocalization invariably leads to large fluctuations in the beads' positions, thereby deteriorating the LBO analysis, the LBO parameters were computed by using the centre of mass of each ring-polymer as the particle position\cite{Rossi}. In addition, in order to guarantee a univocal choice of the nearest neighbors of a given particle we employed a three dimensional Delanuay triangulation. The averaged LBO parameters were then used to define different coarse-grained rotationally invariant parameters that are sensitive to different spatial symmetries\cite{Lechner08}. For example, the invariant $Q_6$ is ideally suited to distinguish between the crystalline and disordered liquid-like configurations, whereas the invariant $Q_4$ allows distinguishing between different types of crystal structures. Conversely, one can construct the invariant $w_6$ that we used to analyze the tendency of the simulated systems to develop icosahedral-like structures \cite{Leocmach12}.

\begin{figure}[t]
\begin{center}
\includegraphics[width=9cm]{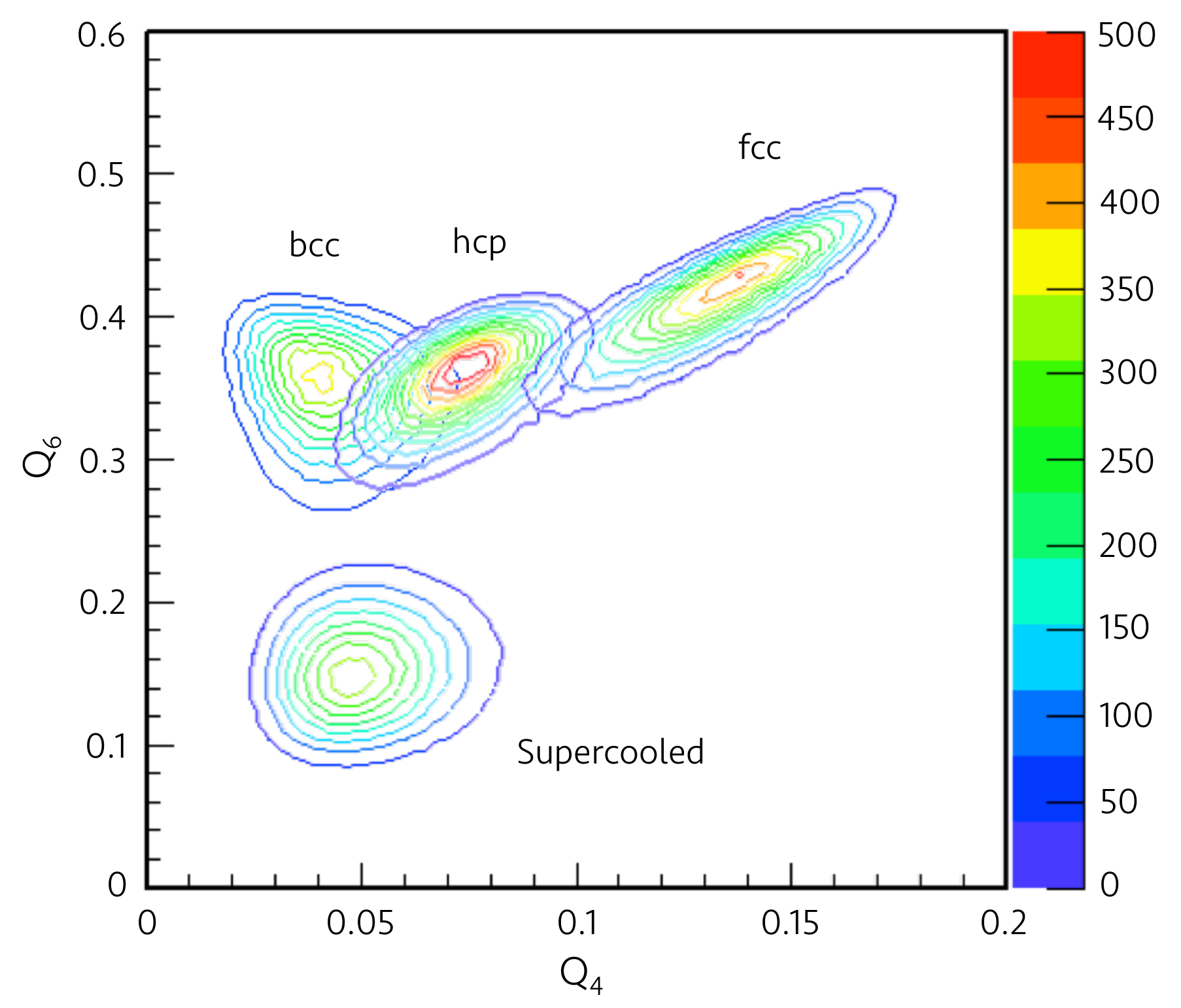}
\end{center}
\caption{Contour plots of the computed probability distributions $p(Q_4,Q_6)$ for pH$_2$ simulated in the bcc, hcp and fcc crystal phases phases at  $T \simeq 12$ K and $\rho=0.0240$ \AA$^{-3}$, and in a metastable liquid state at $T \simeq 13$ K and $\rho=0.0232$ \AA$^{-3}$. Note that the use of periodic boundary conditions compatible with a specific crystal lattice allows simulating also crystalline states different from the equilibrium hcp solid phase of pH$_2$.}
\label{Fig_Q6Q4}
\end{figure}
In order to prevent the crystallization of the mixtures during our PIMC simulations we adopted different strategies. We used a number of particles and a simulation box with side-ratios non compatible with close-packed crystal lattices; moreover, we generated the starting disordered configuration for our simulations as reported in Ref. \onlinecite{Boronat12}. The validity of our choices has been checked by computing the probability distributions $p(Q_4,Q_6)$ for the ordered and disordered configurations, shown in Fig.~\ref{Fig_Q6Q4} for a pure pH$_2$ system\cite{Kuehnel2}; the probability distribution for disordered supercooled liquid pH$_2$ is clearly well separated from the distributions computed for different crystal lattices, indicating the absence of any sign of crystallization in the configurations sampled for the metastable system. Similar results have been obtained for all the quantum binary mixtures considered in the present study.

%


\begin{thebibliography}{99}

\bibitem{Ediger}
M. D. Ediger and P. Harrowell, J. Chem. Phys. \textbf{137}, 080901 (2012). 

\bibitem{Ediger2}
M. D. Ediger, P. Harrowell, and L. Yiu, J. Chem. Phys. \textbf{128}, 034709 (2008).

\bibitem{Nascimento}
M. L. F. Nascimento and E. D. Zanotto, J. Chem. Phys. \textbf{133}, 174701 (2010).

\bibitem{Tang}
C. Tang and P. Harrowell, Nature Mat. \textbf{12}, 507 (2013).

\bibitem{Orava}
J. Orava and A. L. Greer, J. Chem. Phys. \textbf{140}, 214504 (2014).

\bibitem{Wang}
D. Wang, Y. Li, B. B. Sun, M. L. Sui, K. Lu, and E. Ma, App. Phys. Lett. \textbf{84} 4029 (2004).

\bibitem{JRFernandez}
J. R. Fern\'andez and P. Harrowell, Phys. Rev. E \textbf{67}, 011403 (2003).

\bibitem{Coslovich}
D. Coslovich and G. Pastore, J. Chem. Phys. \textbf{127}, 124504 (2007).

\bibitem{Valdes}
L.-C. Valdes, F. Affouard, M. Descamps, and J. Habasaki. J. Chem. Phys. \textbf{130}, 154505 (2009).

\bibitem{Banerjee}
A. Banerjee, S. Chakrabarty, and S. M. Bhattacharyya. J. Chem. Phys. \textbf{139}, 104501 (2013).

\bibitem{Kuehnel}
M. K\"uhnel, J. M. Fern\'andez, G. Tejeda, A. Kalinin, S. Montero, and R. E. Grisenti, Phys. Rev. Lett. \textbf{106} 245301 (2011).

\bibitem{Kuehnel2}
M. K\"uhnel, J. M. Fern\'andez, F. Tramonto, G. Tejeda, E. Moreno, A. Kalinin, S. Montero, D. E. Galli, and R. E. Grisenti, Phys. Rev. B \textbf{89} 180506(R) (2014).

\bibitem{Silvera}
I. F. Silvera, Rev. Mod. Phys. \textbf{52}, 393 (1980).

\bibitem{Challa}
S. R. Challa and J. K. Johnson, J. Chem. Phys. \textbf{111}, 724 (1999).

\bibitem{Fernandez}
J. M. Fern\'andez, M. K\"uhnel, G. Tejeda, A. Kalinin, R. E. Grisenti, and S. Montero, AIP Conf. Proc. {\bf 1501}, 1296 (2012).

\bibitem{Brouwer}
J. P. Brouwer, C. J. N. Van Den Meijdenberg, and J. J. M. Beenakker, Physica \textbf{50} 93 (1970).

\bibitem{Barylnik}
A. S. Baryl'nik, A. I. Prokhvatilov, M. A. Strzhemechny\u{\i}, and G. N. Shcherbakov, Low Temp. Phys. \textbf{19}, 447 (1993).

\bibitem{Galtsov}
N. N. GalÕtsov, A. I. Prokhvatilov, and M. A. Strzhemechny\u{\i}, Low Temp. Phys. \textbf{31}, 947 (2004).

\bibitem{Belan}
V. G. Belan, N. N. GalÕtsov, A. I. Prokhvatilov, and M. A. Strzhemechny\u{\i}, Low Temp. Phys. \textbf{31}, 947 (2005).

\bibitem{Kozioziemsky}
B. J. Kozioziemski and G. W. Collins. Phys. Rev. B \textbf{67}, 174101 (2003).

\bibitem{Kranendonk}
J. V. Kranendonk, \textit{Solid Hydrogen} (Plenum, New York, 1983).

\bibitem{Jackson}
K. A. Jackson, Inter. Sci. \textbf{10}, 159 (2002).

\bibitem{Royall}
C. P. Royall  and S. R. Williams, Phys. Rep. \textbf{560}, 1 (2015).

\bibitem{Stueber}
S. St\"uber, D. Holland-Moritz, T. Unruh, and A. Meyer, Phys. Rev. B \textbf{81} 024204 (2010).

\bibitem{Kuhn}
P. Kuhn, J. Horbach, F. Kargl, A. Meyer, and Th. Voigtmann, Phys. Rev. B \textbf{90} 024309 (2014).

\bibitem{Hopkins}
A. B. Hopkins, F. H. Stillinger, and S. Torquato, Phys. Rev. E \textbf{85} 021130 (2012).

\bibitem{Hirata}
A. Hirata, L. J. Kang, T. Fujita, B. Klumov, K. Matsue, M. Kotani, A. R. Yavari, and M. W. Chen, Science \textbf{341}, 376 (2013).

\bibitem{Silvera2}
I. F. Silvera and V. V. Goldman, J. Chem. Phys. \textbf{69}, 4209 (1978).

\bibitem{Ceperley95}
D. M. Ceperley, Rev. Mod. Phys. \textbf{67},  279 (1995).

\bibitem{Rossi09}
M. Rossi, M. Nava, L. Reatto, and D. E. Galli, J. Chem. Phys. \textbf{131}, 154108 (2009).

\bibitem{Steinhardt83}
P. Steinhardt, D. R. Nelson, and M. Ronchetti, Phys. Rev. B \textbf{28}, 784 (1983).

\bibitem{Lechner08}
W. Lechner, and C. Dellago, J. Chem. Phys. \textbf{129}, 114707 (2008).

\bibitem{Rossi}
M. Rossi, E. Vitali, L. Reatto, and D. E. Galli, Phys. Rev. B \textbf{85}, 014525 (2012).

\bibitem{Leocmach12}
M. Leocmach, and H. Tanaka, Nat. Commun. \textbf{3}, 974 (2012).

\bibitem{Boronat12}
O. N. Osychenko, R. Rota, and J. Boronat, Phys. Rev. B \textbf{85}, 224513 (2012).

\end{thebibliography}
\end{document}